\documentclass[11pt,bm,aps,nofootinbib,amsfonts,amssymb,preprintnumbers]{revtex4}

 \usepackage{here}

\usepackage[dvipdfmx]{graphicx}
\usepackage{amssymb,amsfonts,amsmath,bm,color,multirow,cases,empheq,hyperref}
\usepackage{mathrsfs}

\newcommand{\sech}{\, {\rm sech}}
\usepackage{enumerate}
\usepackage{ulem}
\usepackage{afterpage}

\hypersetup{%
 setpagesize=false,
 bookmarksnumbered=true,%
 bookmarksopen=true,%
 colorlinks=true,%
 linkcolor=blue,%
 citecolor=red}

\begin{document}




\title{Acoustic black and white holes of potential flow in a tube}

\author{ Ren Tsuda}
\email{rtsuda515@g.chuo-u.ac.jp}
\affiliation{Department of Physics, Chuo University, Kasuga, Bunkyo-ku, Tokyo 112-8551, Japan}

\author{Shinya Tomizawa}
\email{tomizawa@toyota-ti.ac.jp}

\author{ Ryotaku Suzuki}
\email{sryotaku@toyota-ti.ac.jp}

\affiliation{Mathematical Physics Laboratory, Toyota Technological Institute\\
Hisakata 2-12-1, Nagoya 468-8511, Japan}
\date{\today}

\preprint{TTI-MATHPHYS-20}




\begin{abstract} 
We propose a new simple model of acoustic black hole in a thin tube, where the difference in the gravitational potential is used to create a transonic flow.
The main merit of our transonic flow model is that the Euler equations can be solved analytically. 
In fact,  we can obtain an exact solution to the equation in terms of a height function in the monatomic case $\gamma=5/3$. For arbitrary $\gamma$, we find that it takes a simple form by the near-sonic approximation.
Moreover, we obtain two analytic solutions describing a backward wave and a forward wave, from which we can confirm the existence of sonic horizons.
\end{abstract}
\date{\today}

\maketitle




\section{Introduction}

\label{sec:intro}

In 1974, Hawking theoretically predicted that a black hole can cause a black body radiation by quantum effects, which is referred to as Hawking radiation~\cite{Hawking1975}.
The Hawking radiation with  a thermal spectrum causes black hole evaporation by pair creation of a particle and an antiparticle in the neighborhood of an event horizon and also yields today's unresolved problem of information loss paradox~\cite{Hawking:1976ra}. 
For black holes in the Universe, the temperature of Hawking radiation is so low that it is considered to be  difficult to observe it.

\medskip

However, the essence of the Hawking radiation does not lie in astrophysical black holes themselves but rather in the spacetime structure of an event  horizon. 
Therefore, it is expected that a similar physical system  may also exhibit  something like Hawking radiation.
From this point of view, in 1981, Unruh demonstrated that the acoustic analogue of black holes admits the thermal spectrum of the Hawking radiation~\cite{Unruh1981}. 
So far, many researchers have proposed various analogue models in different fields of physics. As for the hydrodynamical system, we have analog black holes in the Laval nozzle models~\cite{Sakagami:2001ph,Furuhashi:2006dh,Okuzumi:2007hf,Cuyubamba:2013iwa,daRocha:2017lqj}, surface gravity wave models~\cite{Schutzhold:2002rf}, and draining bathtub models~\cite{Visser:1997ux,basak2003a,  basak2002b,Berti:2004ju,Oliveira2010,Cherubini2005,Cherubini:2011zza}. For other fields, the analog models are also proposed in the Bose-Einstein condensation (BEC)~\cite{Garay:1999sk,Barcelo:2001ca}, electronic wave guide~\cite{Schutzhold:2004tv}, superfluid ${}^3{\rm He}$~\cite{Jacobson:1998ms}, and so on.
Using these analogies, many interesting physics involving the Hawking radiation have been actually observed in the laboratory experiments such as surface gravity waves~\cite{Rousseaux:2007is,
Jannes:2010sa,Weinfurtner:2010nu,
Rousseaux:2010md,
Torres:2016iee,Euve:2021mnj}\footnote{Recently, it was argued that the analogy in the surface gravity wave should be treated carefully since the analogy no longer works for the nonlinear regime~\cite{Euve:2021mnj}, in which some earlier experiments (say Ref.~\cite{Weinfurtner:2010nu}, for example) took place.}, optical fibers~\cite{Philbin:2007ji,Faccio:2009yw}, and
BEC~\cite{Lahav:2009wx,
Steinhauer:2014dra,
MunozdeNova:2018fxv,
Kolobov:2019qfs}
 (also see the reviews~\cite{Barcelo:2005fc,Almeida:2022otk}).
 
 \medskip
 
 For the realization in the hydrodynamics,
a simple one-dimensional model is the Laval nozzle model~\cite{Sakagami:2001ph}, where changing the cross section of the flow creates the sound horizon at the narrow throat of the nozzle.
Another simple model is the draining bathtub model~\cite{Visser:1997ux}, where the sound horizon is formed by the steady planer flow only with radial and tangential velocity.
The bathtub models can be used to prove the superradiant instability that involves a rotating horizon~\cite{basak2003a, basak2002b,Berti:2004ju}.

\medskip

In many previous models of acoustic black hole, contribution of gravity is ignored, and variation of pressure is used to create a transonic flow.
In these models, since the main concern was the sonic horizon and its neighborhood, the global structure was not solved analytically, due to the nonlinearity of the Euler equations. In this article, we consider a new type of acoustic analog model, in which the gravitational potential plays the main role in making the transonic configuration.
In this model, we find the solution of an analog black hole analytically in case of monatomic fluid.
Moreover, we can obtain two approximately solutions in the case that the flow velocity is near the speed of sound.
One is the wave propagating in the same direction as the background flow, and the other is in the different direction.
From the backward wave, one can see that this model has black and white hole horizons.

\medskip

The rest of this article is devoted to the analysis of our new acoustic model.
In the next section, we briefly review the general theory on analog black holes, particularly, the analogy between the Schwarzschild black hole metric written in Painlev\'e-Gullstrand coordinates and  an acoustic metric describing by a perfect fluid.
In Sec.~\ref{sec:setup}, explaining our setup in detail, we propose our new acoustic model, which is a one-dimensional tube model with acoustic black and white holes.
Then, we solve the fluid equations to obtain an exact solution.
 In Sec.~\ref{sec:propagation}, we solve the perturbed equations by a near-sonic approximation.  
In Sec.~\ref{sec:causal},  we discuss the causal structure of the spacetime which the acoustic metric describes by the conformal diagram. 
In Sec.~\ref{sec:summary}, we summarize our results and discuss possible generalization.




\section{Acoustic metric}

\label{sec:metric}

We briefly explain how the acoustic geometry appears from the fluid dynamics, which was first studied by Unruh~\cite{Unruh1981}. 
Let us consider an inviscid perfect fluid that follows the continuity equation
\begin{align}
\partial_0 \rho + \partial_i \left( \rho v^i \right) = 0,
\label{eq:continuity}
\end{align}
and the Euler equations
\begin{align}
\partial_0 v^i + v^j \partial_j v^i = - \frac{1}{\rho}\delta^{ij} \partial_j p + \mu^i,\label{eq:euler}
\end{align}
where $\mu^i$ is an external force per unit mass. We also assume the equation of state for the ideal gas
\begin{align}
\frac{p}{ \rho T} = \mathrm{const.}
\end{align}
With the barotropic condition $p=p(\rho)$ and the adiabatic condition, we also have
\begin{align}
p \rho^{-\gamma} = \mathrm{const},
\label{eq:barotropy}
\end{align}
where $\gamma$ is the heat capacity ratio.
For an irrotational flow, the velocity can be expressed by the velocity potential $\phi$ as
\begin{align}
v^i = - \delta^{ij} \partial_j \phi.
\end{align}
Therefore, the fluid equations~(\ref{eq:continuity}) and (\ref{eq:euler}) reduce to the equations for $\rho$ and $\phi$.

Now we consider a small perturbation around a background flow
\begin{align}
\rho\to \rho_{\rm bg} + \tilde{\rho},\quad \phi \to \phi_{\rm bg} + \tilde{\phi},
\end{align}
which leads to a master equation for $\tilde{\phi}$:
\begin{align}
\label{eq:eqpvp} 
0 = - \left( \partial_0 + \partial_i v_{\rm bg}^i + v_{\rm bg}^i \partial_i \right)
\left[ \frac{\rho_{\rm bg}}{c_\mathrm{s}^2} \left( \partial_0 \tilde{\phi} + v_{\rm bg}^j \partial_j \tilde{\phi} \right) \right]
+ \delta^{ij} \partial_i \left( \rho_{\rm bg} \partial_j \tilde{\phi} \right),
\end{align}
where we denote $ v^i_\mathrm{bg} = - \delta^{ij} \partial_j \phi_\mathrm{bg} $.
Note that $c_\mathrm{s}$ is the speed of sound in the rest frame background fluid given by
\begin{align}
c_\mathrm{s}^2 \left( x \right) = \frac{dp_\mathrm{bg}}{d \rho_\mathrm{bg}} = \frac{\gamma p_\mathrm{bg}}{\rho_\mathrm{bg}}.
\label{eq:sonic}
\end{align}

In fact, this has the same form as the Klein--Gordon equation for a massless scalar field on the following metric, i.e., {\it acoustic metric},
\begin{align}
d s_\mathrm{(ac)}^2 = \frac{  \rho_\mathrm{bg} \left( x \right) }{c_\mathrm{s} \left( x \right) } \left[ 
- \left( c_\mathrm{s}^2 \left( x \right) - v_\mathrm{bg}^2 \left( x \right) \right) dt^2 
- 2 \delta_{ij} v_{\mathrm{bg}}^i\left( x \right) dt dx^j 
+ \delta_{ij} dx^i dx^j
\right] .
\label{eq:metric}
\end{align}
This metric describes the Schwarzschild black hole written in the Painlev\'e--Gullstrand coordinate by setting $v_{\rm bg}^r = - c \sqrt{ r_\mathrm{g} / r}$, $v_\mathrm{bg}^\theta = v_\mathrm{bg}^\varphi = 0$, and $c_\mathrm{s}=c$.




\section{Setup}

\label{sec:setup}

So far, the analog horizons have been studied in various setups with transonic flows.
In particular, for the hydrodynamic analog, the draining bathtub~\cite{Visser:1997ux} and the Laval nozzle~\cite{Sakagami:2001ph} models
have been popular models. However, the effect of the gravitational potential has been ignored in both models. In this article, we rather make use of the gravity to realize a simple model of the transonic flow as in the surface gravity wave model~\cite{Schutzhold:2002rf}.

\begin{figure}
\begin{center}
\includegraphics{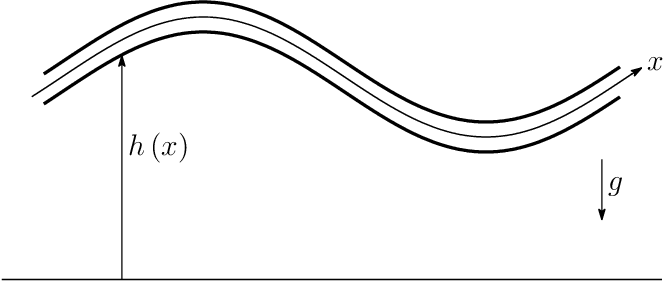}
\caption{Flow in a curved thin tube.}
\label{fig:thin-tube}
\end{center}
\end{figure}

For simplicity, we consider a flow within the thin tube whose height is given by the function $h(x)$ at the coordinate $x$ measured along the tube~(Fig.~\ref{fig:thin-tube}).
Assuming that the tube is thin enough, one can regard the inside flow as one-dimensional flow along $x$. Therefore, the fluid equations (\ref{eq:continuity}), (\ref{eq:euler}) and the wave equation~(\ref{eq:eqpvp}) reduce to
\begin{align}
& \partial_0 \rho +  \partial_x (\rho v) =0
\label{eq:cont-eq-1d},\\
 & \partial_0 v + v \partial_x v = - \frac{1}{\rho} \partial_x p - g \partial_x h,
 \label{eq:euler-eq-1d}
\end{align}
and
\begin{align}
0 = - \left( \partial_0 + \partial_x v + v \partial_x \right)
\left[ \frac{\rho}{c_\mathrm{s}^2} \left( \partial_0 \tilde{\phi} + v \partial_x \tilde{\phi} \right) \right]
+ \partial_x \left( \rho \partial_x \tilde{\phi} \right),
\label{eq:eqpvp-1d}
\end{align}
where $v$ and $g$ are the velocity in the $x$ direction and the gravitational acceleration, respectively. 
$p$ is given by Eq.~(\ref{eq:barotropy}). 
Note that from here on we omit the subscript ``bg" in the background variables.

For the stationary configuration,
the continuity equation~(\ref{eq:cont-eq-1d}) and Poisson relation~(\ref{eq:barotropy}) lead to
\begin{align}
&\rho \left( x \right) v(x)  = \pm \hat{\rho}\, \hat{c}_\mathrm{s} , \\
&c_\mathrm{s}^2 \left( x \right) = \frac{\gamma p(x)}{\rho(x)}= \left( \frac{ \hat{c}_\mathrm{s} }{ \left| v \left( x \right) \right| } \right)^{\gamma +1} v^2 \left( x \right),
\end{align}
where the flow should be either in positive or negative directions. 
Without loss of generality, we can choose $v(x)>0$. 
The constant $\hat{\rho}$ and $\hat{c}_\mathrm{s}$ are determined by the values at the sonic point $x=x_\mathrm{s}$
\begin{align}
\hat{\rho} \equiv \rho \left( x_\mathrm{s} \right),
\quad \hat{c}_\mathrm{s} \equiv v \left( x_\mathrm{s} \right) = c_\mathrm{s} \left( x_\mathrm{s} \right).
\end{align}

\begin{figure}
\begin{center}
\includegraphics[width=8cm]{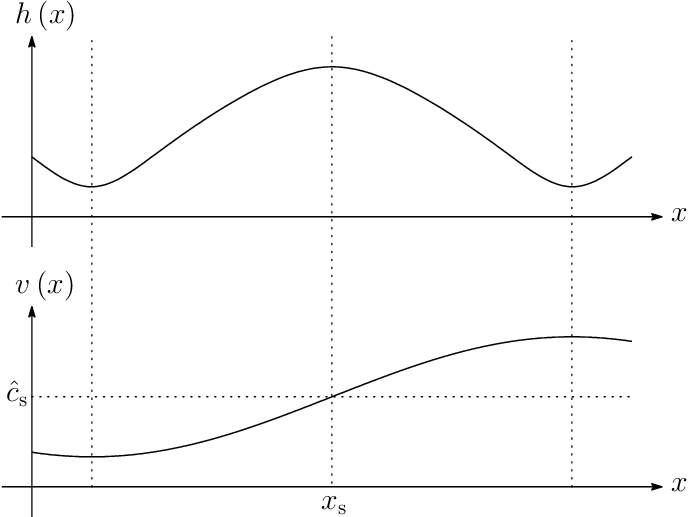}
\caption{Correspondence between the height and velocity profile.} \label{fig:hx-vx}
\end{center}
\end{figure}

For instance, let us consider the height profile as depicted in the upper graph of Fig.~\ref{fig:hx-vx}.
One can see that the detail of the transonic flow is strongly restricted (lower graph of Fig.~\ref{fig:hx-vx}).
The Euler equation~(\ref{eq:euler-eq-1d}) is rewritten as an ODE of $v(x)$
\begin{align}
g h' \left( x \right) = - \left[1-\left( \frac{ \hat{c}_\mathrm{s} }{ v \left( x \right) } \right)^{\gamma+1} \right] v \left( x \right) v' \left( x \right),
\label{eq:eulereq-ode}
\end{align}
where $'$ denotes the derivative with respect to $x$.
This immediately indicates that the sonic points, if they exist, only appear at the stationary points $h'(x)=0$.
By differentiating it, we also have
\begin{align}
g h'' \left( x_\mathrm{s} \right) = - \left( \gamma+1 \right) \left[v' \left( x_\mathrm{s} \right)\right]^2 < 0.
\end{align}
Hence, the sonic points must appear at the local maxima of the potential.
In the similar way, one can easily show the local extrema of the velocity profile correspond to the local minima of the potential.

Equation~(\ref{eq:eulereq-ode}) is integrated to give the Bernoulli formula
\begin{align}
g h \left( x \right)
+  \frac{1}{\gamma-1} \frac{ \hat{c}_\mathrm{s}^{\gamma+1}}{v^{\gamma-1} \left( x \right)}+\frac{v^2 \left( x \right)}{2} =C.
\label{eq:euler-eq-1d-int}
\end{align}
To obtain a transonic flow around a sonic point $x=x_\mathrm{s}$, we set 
\begin{align}
 C= \frac{\gamma+1}{\gamma-1}\frac{\hat{c}_\mathrm{s}^2}{2} + g h(x_\mathrm{s}).
\end{align}
With this, Eq.~(\ref{eq:euler-eq-1d-int}) is rewritten in a dimensionless form
\begin{align}
-\bar{h}(x)
+  \frac{\bar{v}^{1-\gamma}(x)-1}{\gamma-1}+\frac{\bar{v}^2(x)-1}{2}=0,
\label{eq:euler-eq-1d-int2}
\end{align}
where we have defined the normalized velocity and potential as
\begin{align}
\bar{v}(x) \equiv \hat{c}_\mathrm{s}^{-1} v \left( x \right) , 
\quad \bar{h} \left( x \right) \equiv \hat{c}_\mathrm{s}^{-2} g \left( h \left( x_\mathrm{s} \right) - h \left( x \right) \right).
\end{align}
The behavior around the sonic point is determined by expanding $\bar{v} \left( x \right) = 1 + \delta \bar{v} \left( x \right)$,
\begin{align}
\delta \bar{v} \left( x \right) \simeq \pm \sqrt{\frac{2}{\gamma+1}} \sqrt{\bar{h} \left( x \right)}.
\end{align}
Since $\bar{h}(x) \simeq \bar{h}'' \left( x_\mathrm{s} \right) \left( x_\mathrm{s} -x \right)^2$ for $x\simeq x_\mathrm{s}$, 
the smoothness of $\delta \bar{v}(x)$ requires
\begin{align}
 \delta \bar{v} \left( x \right) \simeq \sqrt{\frac{2}{\gamma+1}} \times 
 \left\{ 
 \begin{array}{cc} 
 \displaystyle \mp \sqrt{\bar{h}(x)} & (x\leq x_\mathrm{s})\\
 \displaystyle \pm \sqrt{\bar{h}(x)} & (x>x_\mathrm{s})
 \end{array}
 \right. .
 \label{eq:smoothness}
\end{align}
We could not find the global solution for Eq.~(\ref{eq:euler-eq-1d-int2})  in general, but for the monatomic case ($\gamma=5/3$), we find an analytic solution 
\begin{align}
\bar{v} \left( x \right) =&  
\sqrt{
- 2 \sqrt{\cosh^{3} \frac{\psi}{3}}
+ 3 \sqrt{\cosh \psi}
+3\sqrt{3} \sinh \frac{\psi}{3} \left(  3 \cosh \frac{\psi}{3}
   +\cosh \psi \left(  \sqrt{ \cosh \psi {\rm sech}^3 \frac{\psi}{3}}-2 \right) \right)^{-1/2}
}, \label{eq:transonic-sol}
\end{align}
with
\begin{align}
 \psi =\pm  {\rm arccosh}\left[\left(1+\frac{1}{2}\bar{h}(x)\right)^2\right].
\end{align}
For $x\simeq x_\mathrm{s}$, we have 
\begin{align}
 \bar{v} \left( x \right) \simeq 1 + \frac{\sqrt{3} \psi}{2\sqrt{2}} ,\quad \psi \simeq \pm \sqrt{2\bar{h}}.
\end{align}
Therefore, for the smoothness at $x=x_\mathrm{s}$~(\ref{eq:smoothness}), we must choose
\begin{align}
 \psi=   \left\{ 
 \begin{array}{cc} 
\displaystyle \mp {\rm arccosh}\left[\left(1+\frac{1}{2} \bar{h}(x)\right)^2 \right] & (x\leq x_\mathrm{s})\vspace{1mm}\\
\displaystyle \pm {\rm arccosh}\left[\left(1+\frac{1}{2}\bar{h}(x)\right)^2 \right] & (x>x_\mathrm{s})
 \end{array} \right. .
 \label{eq:smoothness-psi}
\end{align}
This gives the transonic solution for a given height function $h(x)$.




\section{Near-sonic approximation and wave propagation}

\label{sec:propagation}

In this section, we solve the wave propagation (\ref{eq:eqpvp-1d}) in the acoustic geometry derived in the previous section. 
Since it is difficult to obtain exact solutions in general, we consider two simple configurations.




\subsection{Wavy toroidal tube}


\begin{figure}[t]
\begin{center}
\scalebox{1}{\includegraphics{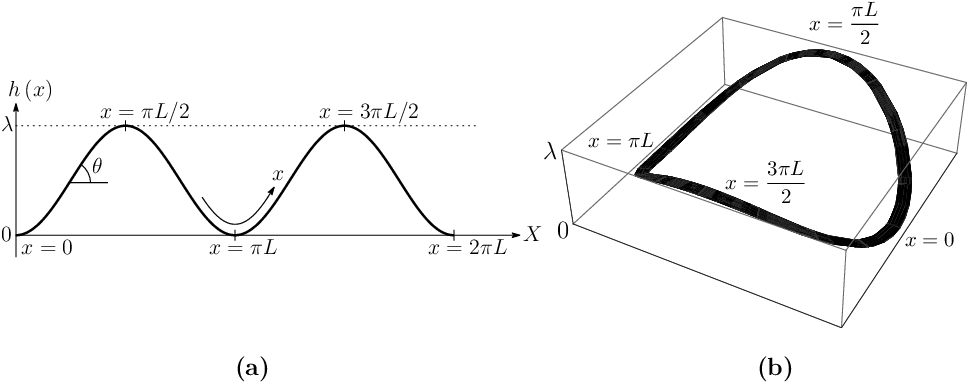}}
\caption{(a) Bent tube of the length $2 \pi L$ in accordance with $ h \left( x \right) $ and (b) wavy toroidal tube. }
\label{fig:model-pics}
\end{center}
\end{figure}

\begin{figure}[t]
\begin{center}
\scalebox{1.0}{\includegraphics{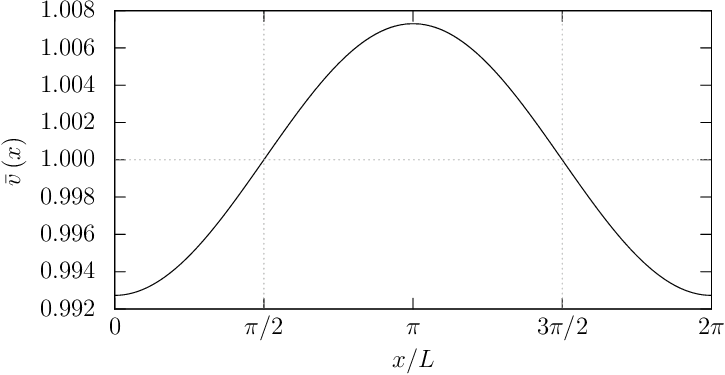}}
\end{center}
\caption{Velocity profile of transonic flow solution in Eq.~(\ref{eq:transonic-sol}).}
\label{fig:transonic-sol}
\end{figure}

First, we consider a periodic model of the length $2\pi L$, whose height is given by the following profile (Fig.~\ref{fig:model-pics}):
\begin{align}
h \left( x \right) = \lambda \sin^2 \frac{x}{L},
\label{eq:hprofile}
\end{align}
where $\lambda$ is the amplitude of the elevation.\footnote{
For the actual implementation, it would be useful to clarify the dependence on the base coordinate $X$. 
From $dX = \sqrt{1-h'(x)^2}dx$, we have
\begin{align}
X = \frac{L}{2} E\left(\frac{2x}{L},\frac{\lambda^2}{L^2}\right),
\quad 
0\leq X \leq 4 L E\left(\frac{\lambda^2}{L^2}\right),
\end{align}
where $E(\phi,k)$ and $E(k)$ are the incomplete and complete elliptic integrals of the second kind, respectively.}
The inclination angle $\theta$ of the tube is given by
\begin{align}
  \sin \theta = h'(x) = \frac{\lambda}{L} \sin \frac{2x}{L},
\end{align}
which restricts the range of $\lambda$ for $0 \leq \lambda \leq L $.
Since the sonic points must be at $x=\pi L/2,3\pi L/2$, the transonic solution for $\gamma=5/3$ is given by Eq.~(\ref{eq:transonic-sol}) with
\begin{align}
 \psi=    \left\{ 
 \begin{array}{cc} 
 \displaystyle \mp {\rm arccosh}\left[\left(1+\frac{1}{2}\bar{\lambda}\cos^2\frac{x}{L}\right)^2\right] & (0 \leq x\leq \pi L/2,3\pi L/2\leq x< 2\pi L)\vspace{1mm}\\
\displaystyle \pm {\rm arccosh}\left[\left(1+\frac{1}{2}\bar{\lambda}\cos^2\frac{x}{L}\right)^2\right] & (\pi L/2 < x <3 \pi L/2)
 \end{array}
 \right. ,
 \label{eq:smoothness-psi-1}
\end{align}
where 
\begin{align}
\bar{\lambda} \equiv \frac{g \lambda}{ \hat{c}_\mathrm{s}^2 },\label{eq:def-barlam}
\end{align}
and the sign is set so that it flips at each sonic points.
In Fig.~\ref{fig:transonic-sol}, we show a typical velocity profile of the transonic solution, 
where we assumed a laboratory-sized system with $L=0.5~\mathrm{m}$ and $\lambda=5\times10^{-2}~\mathrm{m}$, 
and superfluid $\rm {}^4 He$ ($M_{\rm {}^4He}=4.0026\times 10^{-3}~{\rm kg/mol}$, $\gamma=5/3$) 
at the sonic points temperature of $\hat{T}=2~{\rm K}$, 
also with $g=9.798~{\rm m/s^2}$ and $R=8.314~{\rm m^2~kg~s^{-2}~K^{-1}~mol^{-1}}$.
The value of $\bar{\lambda}$ is given by $\bar{\lambda} = 7.075\times 10^{-5}$.

To solve the wave equation~(\ref{eq:eqpvp-1d}) analytically, we further assume the small amplitude of the elevation $\bar{\lambda} \ll 1$ where the flow becomes almost sonic. As shown above, this assumption is rather realistic in the laboratory experiment.
The small undulation allows us, as a by-product, to obtain the analytic form of the velocity profile for general $\gamma$ as in Eq.~(\ref{eq:smoothness}),
\begin{align}
\bar{v} \left( x \right) = 1
- \sqrt{ \frac{ 2 \bar{\lambda} }{ \gamma + 1 } } \cos \frac{x}{L} 
+ {\cal O} \left( \bar{\lambda} \right)
\quad 
\mbox{for}
\quad 
\bar{\lambda} \ll 1.
\end{align}
In the same way, the sonic speed $c_\mathrm{s}$ and the mass density $\rho$
are expanded as
\begin{align}
c_\mathrm{s} \left( x \right) =& \hat{c}_\mathrm{s} 
+ \hat{c}_\mathrm{s}(\gamma-1) \sqrt{ \frac{  \bar{\lambda} }{ 2(\gamma + 1 )} } \cos \frac{x}{L}
+ {\cal O} \left( \bar{\lambda} \right), \\
\rho \left( x \right) =& \hat{\rho} 
+ \hat{\rho} \sqrt{ \frac{ 2 \bar{\lambda} }{ \gamma + 1 } } \cos \frac{x}{L}
+ {\cal O} \left( \bar{\lambda} \right).
\end{align}
We will refer to this formulation as the {\it near-sonic approximation}.
With these backgrounds, Eq.~(\ref{eq:eqpvp-1d}) can be solved by expanding in $\displaystyle \sqrt{\bar{\lambda}}$.

First, we begin with the waves at the limit $\bar{\lambda} \to 0$, where the wave equation  (\ref{eq:eqpvp-1d}) reduces to
\begin{align}
0 = \partial_0 \left( \partial_0 \tilde{\phi} + 2 \hat{c}_\mathrm{s} \partial_x \tilde{\phi} \right),
\end{align}
which has solutions
\begin{align}
\label{eq:0th-sol}
\tilde{\phi} = \exp \left[ - i \left( \omega t - k x \right) \right],
\quad \omega = 0 ,\ 2k \hat{c}_\mathrm{s} .
\end{align}
The wave with $\omega = 2 k \hat{c}_\mathrm{s}$ corresponds to the forward wave that propagates in the same direction as the background sonic flow,
and the one with $ \omega = 0 $, the backward wave that tries to go back against the background.

To study the transonic effect, we assume following wave form:
\begin{align}
 \tilde{\phi}(t,x) = \exp\left[-i \omega t + i \Psi(x)\right].
\end{align}
For the forward wave, assuming $\omega={\cal O}(1)$, we can expand the phase as
\begin{align}
\Psi^{\rm (for)}(x) = k_0 x + \sqrt{\lambda}\,   \psi_1^{(\rm for)}(x)+{\cal O}(\lambda),\quad k_0 := \omega/(2\hat{c}_{\rm s}).
\end{align}
By expanding Eq.~(\ref{eq:eqpvp-1d}), we obtain
\begin{align}
0=& \sqrt{\frac{2}{\gamma+1}} \left( \frac{ L \omega }{\hat{c}_\mathrm{s}} \cos \frac{x}{L} - 2 i \sin \frac{x}{L} \right)
- \frac{8 L}{3-\gamma} \partial_x   \psi_1^\mathrm{(for)}(x) .\label{eq:1/2th-forwardeq}
\end{align}
This can be solved as
\begin{align}
 \psi_1^{(\rm for)}(x) = {\rm const.}+ \frac{3-\gamma}{8} \sqrt{\frac{2}{1+\gamma}}\left( \frac{L\omega}{\hat{c}_{\rm s}} \sin\frac{x}{L} + 2 i \cos\frac{x}{L} \right).\label{eq:1/2th-forwardeq-sol}
\end{align}
Therefore, the forward wave solution is given by
\begin{align}
\tilde{\phi}^{(\rm for)}(t,x) = A^{(\rm for)}(x) \exp\left[ -i \omega t + i\tilde{\Psi}^{\rm (for)}(x) \right],\label{eq:forward-sol-res}
\end{align}
where we rewrote the modulations in the amplitude and phase separately as
\begin{align}
&A^{(\rm for)}(x) = \exp\left( -\frac{3-\gamma}{4} \sqrt{\frac{2\bar{\lambda}}{1+\gamma}}\cos \frac{x}{L} \right),\\
&\tilde{\Psi}^{(\rm for)}(x) = \frac{\omega}{2\hat{c}_{\rm s}} x + \frac{3-\gamma}{8}\sqrt{\frac{2\bar{\lambda}}{1+\gamma}} \frac{L\omega}{\hat{c}_s} \sin \frac{x}{L}.
\end{align}
The local phase velocity is then also expanded by
$\displaystyle{\sqrt{\bar{\lambda}}}$
\begin{align}
 v^{(\rm for)}_{\rm ph}(x) = \frac{\omega}{d\tilde{\Psi}^{(\rm for)}(x)/dx} = 2 \hat{c}_{\rm s} \left(1-\frac{3-\gamma}{4} \sqrt{\frac{2\bar{\lambda}}{1+\gamma}} \cos\frac{x}{L}\right).
\end{align}

\begin{figure}[t]
\begin{center}
\scalebox{.95}{\includegraphics{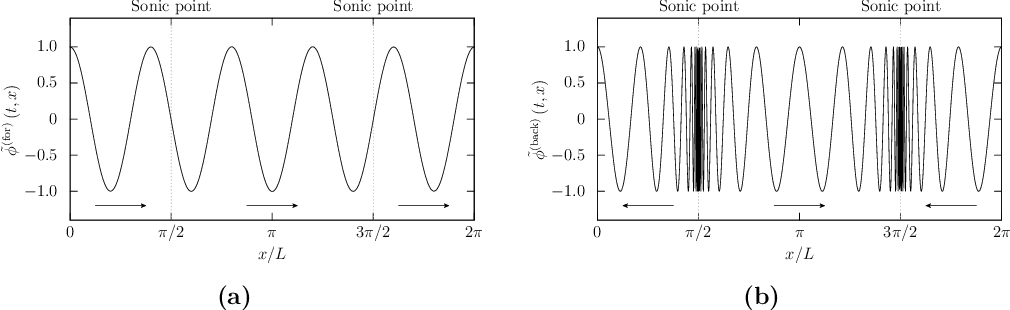}}
\caption{Plots of the waves in the case of $\bar{\lambda} = 7.075\times 10^{-5}$ and $\gamma=5/3$. 
(a) The forward going wave with $\omega = 10 \hat{c}_\mathrm{s} / L $, and (b) the backward going one with $ \omega_1 = 10 \hat{c}_\mathrm{s} / L $.\label{fig:waves}}
\end{center}
\end{figure}

For the backward wave, we should start with rescaling $\omega$ as 
\begin{align}
 \omega = \omega_1 \sqrt{\bar{\lambda}},
\end{align}
since it gives  $\omega=0$ at the limit $\bar{\lambda}\to 0$.
Then, it turns out the mode function should satisfy
\begin{align}
0= \left( \sqrt{\frac{2}{\gamma + 1}} \frac{i  \omega_1}{\hat{c}_{\rm s}}  \sec\frac{x}{L}
- L^{-1}\tan \frac{x}{L} \right) \partial_x e^{i\Psi^\mathrm{(back)}(x)}  +  \partial_x \partial_x e^{i\Psi^\mathrm{(back)}(x)},
\end{align}
which is solved as
\begin{align}
 e^{i\Psi^\mathrm{(back)}(x)} = C_0 + C_1 \left(\frac{1-\sin\frac{x}{L}}{1+\sin\frac{x}{L}}\right)^\frac{i L \omega_1}{\hat{c}_{\rm s}\sqrt{2(1+\gamma)}}.
\end{align}
Since we are interested in the propagating solution, we choose $C_0=0$ and $C_1=1$, which lead to
\begin{align}
 \Psi^\mathrm{(back)}(x) = \frac{L\omega_1}{\hat{c}_{\rm s}\sqrt{2(1+\gamma)}} \log \left(\frac{1-\sin\frac{x}{L}}{1+\sin\frac{x}{L}}\right).
\end{align}
This wave actually propagates backward in the subsonic region and forward for the supersonic region
\begin{align}
 v^{(\rm back)}_{\rm ph}(x) = \frac{\omega}{d\Psi^{(\rm back)}(x)/dx} = -\hat{c}_{\rm s}\sqrt{\frac{(\gamma+1)\lambda}{2}}\cos\frac{x}{L}.
\end{align}
In Fig.~\ref{fig:waves} we give plots of these forward and backward waves.

\subsection{Infinite tube with a single bump}

\begin{figure}[t]
\begin{center}
\includegraphics[width=8cm]{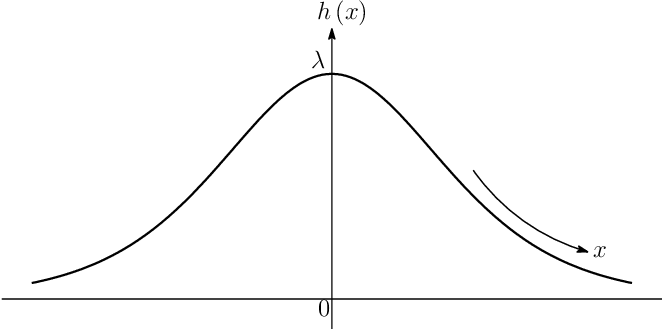}
\caption{Infinite tube with a single bump.\label{fig:singlebump}}
\end{center}
\end{figure}

\begin{figure}[t]
\begin{center}
\includegraphics[width=16cm]{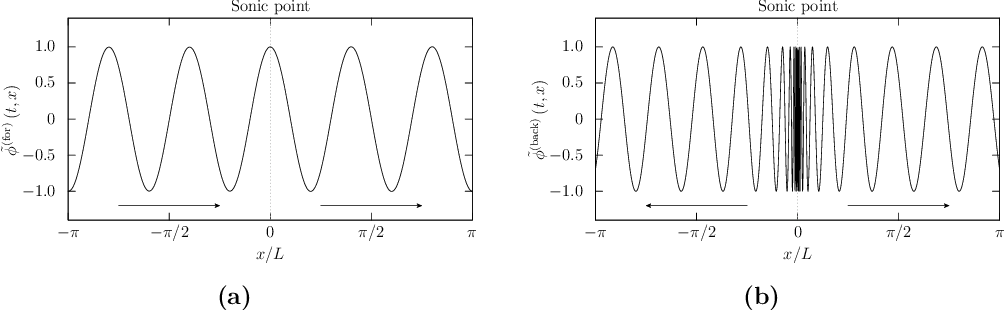}
\caption{Plots of (a) the forward going and (b) backward going waves with the same parameter choice as Fig.~\ref{fig:waves}.
\label{fig:sb-waves}}
\end{center}
\end{figure}

Another simple situation is a tube with the infinite length with a single bump (Fig.~\ref{fig:singlebump})
\begin{align}
 h(x) = \lambda \sech^2 \frac{x}{L},
\end{align}
where $L$ gives the width of the bump, and the amplitude $\lambda$ is limited in the range of $0\leq \lambda \leq (3\sqrt{3}/4)L$ as in the toroidal model.
The sonic point only exists at $x=0$. 
For the small elevation, Eq.~(\ref{eq:smoothness}) leads to
\begin{align}
 \bar{v}(x) = 1 + \sqrt{\frac{2\bar{\lambda}}{\gamma+1}} \tanh\frac{x}{L} + {\cal O}(\bar{\lambda}),
\end{align}
where we set the sign so that $x<0$ is subsonic and $x>0$ supersonic, and $\bar{\lambda}$ is defined in the same way as Eq.~(\ref{eq:def-barlam}).
With the same analysis as in the toroidal tube, we obtain the forward wave~(\ref{eq:forward-sol-res}) but with
\begin{align}
&A^{(\rm for)}(x) = \exp\left( \frac{3-\gamma}{4} \sqrt{\frac{2\bar{\lambda}}{1+\gamma}}\tanh\frac{x}{L} \right),\\
&\tilde{\Psi}^{(\rm for)}(x) = \frac{\omega}{2\hat{c}_{\rm s}} x - \frac{3-\gamma}{8}\sqrt{\frac{2\bar{\lambda}}{1+\gamma}} \frac{L\omega}{\hat{c}_s} \log \cosh\frac{x}{L},
\end{align}
where the phase velocity becomes
\begin{align}
v^{\rm (for)}_{\rm ph}(x) = 2 \hat{c}_{\rm s} + \frac{3-\gamma}{\sqrt{2(1+\gamma)}}\hat{c}_{\rm s}\sqrt{\bar{\lambda}} \tanh\frac{x}{L}.
\end{align}
The backward wave is given by
\begin{align}
\tilde{\phi}^{\rm (back)}(t,x) = e^{-i \sqrt{\bar{\lambda}} \omega_1 t}  \left|\sinh\frac{x}{L}\right|^{\frac{i L \omega_1}{\hat{c}_{\rm s} } \sqrt{  \frac{2}{\gamma+1}}},
\end{align}
where the phase velocity becomes
\begin{align}
 v^{(\rm back)}_{\rm ph}(x) = \hat{c}_{\rm s} \sqrt{\frac{1+\gamma}{2}} \sqrt{\bar{\lambda}} \tanh\frac{x}{L}.
\end{align}
The two waves are shown in Fig.~\ref{fig:sb-waves}.

\section{Acoustic metric and causal structure}

\label{sec:causal}

\begin{figure}[h]
\begin{center}
\includegraphics[width=6cm]{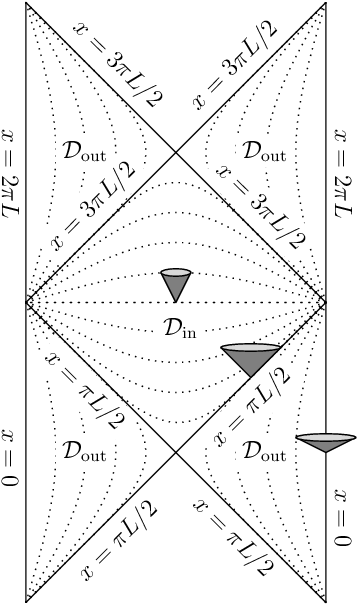}
\caption{Conformal diagram of an acoustic black and white hole spacetime.
The sonic points $x= \pi L/2 , ~ t= \infty$ and $x=3 \pi L/2 , ~ t=\infty$ act as black hole horizon ${\cal H}_{\rm B}$ and white hole horizon ${\cal H}_{\rm W}$, respectively.
One should note that the sonic velocity $c_\mathrm{s}(x)$ depends on $x$, and hence, the sound cones differ.}
\label{fig:conformal}
\end{center}
\end{figure}

Finally, we discuss that the causal structure of the spacetime corresponds to the toroidal model.
The acoustic metric is written as Eq.~(\ref{eq:metric}).
In order to see the causal structure of the spacetime described by the acoustic metric, 
let us consider the conformally transformed metric
\begin{align}
d \tilde s_\mathrm{(ac)}^2 = 
- \left( c_\mathrm{s}^2 \left( x \right) - v^2 \left( x \right) \right) dt^2 
- 2 v \left( x \right) dt dx 
+ dx^2 + dy^2 + dz^2, 
\end{align}
which can be written as
\begin{eqnarray}
d \tilde s_\mathrm{(ac)}^2 =-\left(1-\frac{v^2(x)}{c_s^2(x)} \right)c_\mathrm{s}^2(x)dt_*^2
+\frac{1}{\displaystyle 1-\frac{v^2(x)}{c_\mathrm{s}^2(x)}}dx^2+ dy^2 + dz^2, \label{eq:Sch}
\end{eqnarray}
where
\begin{eqnarray}
dt_*=dt+\frac{v(x)}{c_\mathrm{s}^2(x)-v^2(x)}dx.
\end{eqnarray}
The two-dimensional  $(t_*,x)$  part of the metric is similar to the  Schwarzschild metric, where  one must note that  the radial coordinate $x$, unlike the Schwarzschild metric, has the finite range $0\le x\le 2\pi L$, so the spacetime has no infinity.
Therefore, the Schwarzschild-like metric~(\ref{eq:Sch}) describes the spacetime displayed by the conformal diagram in Fig.\ref{fig:conformal}, which consists of the following four portions.

\medskip

\begin{itemize}
\item An outer region of black and white holes : ${\cal D_{\rm out}}=\{ (t,x)\ | \  -\infty<t<\infty,\ 0\leq x < \frac{\pi }{2}L,\ \frac{3\pi }{2}L< x \leq 2\pi L\}$. 
In this region, the background velocity is smaller than the sonic velocity, i.e.,  $v(x)<c_\mathrm{s}(x)$.
\item A black hole horizon : ${\cal H_{\rm B}}=\{ (t,x)\ | t=\infty,\  x=\frac{\pi}{2}L\}$:  In this point, the background velocity coincides with the sonic velocity $v(x)=c_\mathrm{s}(x)$. 
\item An inner region of black and white holes : ${\cal D_{\rm in}}=\{ (t,x)\ | \  -\infty<t<\infty,\ \frac{\pi }{2}L<x< \frac{3\pi }{2}L\}$. 
In this region, the background velocity is larger than the sonic velocity, i.e., $v(x)>c_\mathrm{s}(x)$
\item A white hole horizon : ${\cal H_{\rm W}}=\{ (t,x)\ | t=\infty,\  x=\frac{3\pi}{2}L\}$:  In this point, the background velocity coincides with the sonic velocity $v(x)=c_\mathrm{s}(x)$. 
\end{itemize}




\section{Summary and discussions}

\label{sec:summary}

We have proposed a simple model of the acoustic black hole.
Unlike many conventional models which make use of a pressure difference, 
our model avails itself of the gravitational potential to create a transonic flow.
The main advantage of our model is that we can solve the Euler equations analytically.
In particular, we have found an exact solution to the equations
for a given height function in the monatomic case of $\gamma=5/3$.
Under the near-sonic approximation, we obtain two wave solutions in the toroidal tube and infinite tube setups:
One describes a wave propagating in the same direction as the background flow, 
and the other the backward going one.
One can see from the backward wave solution that there are two sonic horizons, black and white hole horizons,
in our periodic transonic flow model.

\medskip

Of course, there are some shortcomings in our model. 
The toroidal bending of the tube causes the centrifugal force on the fluid, whose effect is ignored in the analysis.
The inertia force for a fluid element per unit mass is estimated as $\hat{c}_\mathrm{s}^2/R$ where $\hat{c}_\mathrm{s}$ is the sonic velocity and $R$ is the curvature radius of the bending tube. 
Since we have $ 10^2 \lesssim \hat{c}_\mathrm{s} \lesssim 10^3~\rm{m/s}$ for ordinary fluids, 
this rather dominates over the gravitational acceleration $g \sim 10~\rm{m/s^2}$ in the laboratory experiment. Nevertheless, the inertia would not matter in the thin tube approximation since the tube provides the supporting force.  
However, with a finite cross section, the strong inertia would make strong inhomogeneity in the density and pressure profiles for given $x$, which will affect the wave propagation as the finite size effect.
To incorporate this, one has to start from the three-dimensional system and reduce it to that of one dimension.
The lack of energy supply is another unphysical assumption. To make a similar steady flow in the laboratory, one has to place a pump to compensate the dissipation.
The pump will be placed in the middle of the flow on either of subsonic or supersonic sides, which will cut the circular topology of the analog spacetime.
For these reasons, our toroidal model will not be considered physically reasonable, but we believe that it is still considered pedagogically worthwhile as a toy model for learning about properties of an acoustic black hole.

\medskip

In general, one  of important advantages to consider analog models is that one can carry out an experiment with respect to black hole physics in a laboratory. 
However, to do so, at least, in the fluid system, the sonic horizon must be stable since 
a shock wave may appear near the horizon.
At present,  we do not know the stability in our model, which deserves our future works.
In addition, we also should see whether the wave propagation  has the similar nature to that of Hawking radiation. 
Moreover, we can also consider various generalizations of our model if we give  appropriate forms of the  function $h(x)$. 
For instance, if we replace  $x/L$ in $h(x)$ with $nx/L$, we can construct a multiblack hole system of $n$ black and white holes. 
Furthermore, if we choose the function $h(x)$ with periodicity $2\pi L$ such that $h(0)=h(2\pi L)=0$, $h'(\pi L/2)=h'(\pi 3L/2)=0$ and $h''(\pi L/2)=0$, we can consider the analog model of black hole binary.

\acknowledgments
The authors thank Kouji Nakamura and Ken-ichi Nakao for useful comments and discussion during the 23rd Singularity meeting.
The authors also thank Germain Rousseaux for kind comments on the literature.
This work is supported by Toyota Technological Institute Fund for Research Promotion A. 
R. S. was supported by JSPS KAKENHI Grant No. JP18K13541. 
S. T. was supported by JSPS KAKENHI Grant Nos. 21K03560 and17K05452.




\end{document}